**Grains and grain boundaries in highly crystalline monolayer molybdenum disulfide**

Arend M. van der Zande*#1,2, Pinshane Y. Huang*3, Daniel A. Chenet*2, Timothy C. Berkelbach4, Youmeng You5, Gwan-Hyoung Lee2,6, Tony F. Heinz1,5, David R. Reichman1,4, David A. Muller3,7, James C. Hone1,2

* These authors contributed equally to this work.
# Corresponding Author

Affiliations:

1. Energy Frontier Research Center, Columbia University, New York, New York, 10027, USA

2. Department of Mechanical Engineering, Columbia University, New York, New York, 10027, USA

3. School of Applied and Engineering Physics, Cornell University, Ithaca, New York, 14853, USA

4. Department of Chemistry, Columbia University, New York, New York, 10027, USA

5. Department of Physics and Electrical Engineering, Columbia University, New York, New York, 10027, USA

6. Samsung-SKKU Graphene Center (SSGC), Suwon, 440-746, Korea

6. Kavli Institute at Cornell for Nanoscale Science, Ithaca, New York 14853, USA

**Abstract:**

Recent progress in large-area synthesis of monolayer molybdenum disulfide, a new two-dimensional direct-bandgap semiconductor, is paving the way for applications in atomically thin electronics. Little is known, however, about the microstructure of this material. Here we have refined chemical vapor deposition synthesis to grow highly crystalline islands of monolayer molybdenum disulfide up to 120 μm in size with optical and electrical properties comparable or superior to exfoliated samples. Using transmission electron microscopy, we correlate lattice orientation, edge morphology, and crystallinity with island shape to demonstrate that triangular islands are single crystals. The crystals merge to form faceted tilt and mirror boundaries that are stitched together by lines of 8- and 4- membered rings. Density functional theory reveals localized mid-gap states arising from these 8-4 defects. We find that mirror boundaries cause strong photoluminescence quenching while tilt boundaries cause strong enhancement. In contrast, the boundaries only slightly increase the measured in-plane electrical conductivity.

**Introduction (3000 words total, 500 for introduction):**

Characterizing the structure and properties of grains and grain boundaries is critical for understanding and controlling material properties in the expanding array of two-dimensional (2D) materials[1]. Studies have mapped the grain structure of large-area graphene grown by chemical vapor deposition (CVD), characterized defects such as grain boundaries at the atomic scale, and demonstrated that these grain boundaries can strongly affect graphene's electrical optical and mechanical properties.[2-11] Much less is known about the grain structure and properties of defects in other 2D materials, such as molybdenum disulfide ($MoS_2$). $MoS_2$ has been widely explored in the form of nanotubes[12], nanoparticles[13], and thin films[14-17] owing to its excellent tribological and catalytic properties. Recent work has shown that monolayer $MoS_2$ produced by mechanical exfoliation is a 2D direct band-gap semiconductor[18,19]. Moreover, polycrystalline monolayer $MoS_2$ can now be synthesized in large area using CVD[20-23], making it a promising candidate for building atomically-thin layered electrical[24-26], optical[18,27], and photovoltaic[28] devices. Here, we grow high quality crystals of monolayers $MoS_2$ exhibiting grain size up to 120 μm and possessing optical and electronic properties comparable or superior to those of exfoliated samples. Our study applies a diverse set of characterization methods, combining transmission electron microscopy (TEM) of atomic and crystal structure with optical spectroscopy and electrical transport to determine the influence of grain boundaries on the optical and electrical properties of the material. Using diffraction-filtered TEM imaging, we confirm that the islands are single-domain crystals with an orientation and edge structure that is highly correlated to the island shape. In addition, we characterize the two most common types of grain boundaries present in these films: tilt grain boundaries, which mainly occur where two nucleation islands intersect, and mirror twin boundaries, which occur in single islands. Using atomic-resolution imaging, we determine that the grain boundaries are stitched together predominantly through lines of 8 and 4 membered rings, and we model the electronic structure of the boundary to show that the defects lead to the appearance of mid-gap states. By correlating grain imaging with photoluminescence and electrical transport measurements on individual grain boundaries, we show that these grain boundaries strongly affect the photoluminescence observed from the $MoS_2$ monolayers, but only slightly increase their in-plane electrical conductivity.

Figure 1(a) is an optical image of a triangular $MoS_2$ island grown on a silicon substrate with an insulating silicon oxide epilayer. We grew the crystals using chemical vapor deposition with solid precursors[20] by placing the substrate in a tube furnace in close proximity to $MoO_3$ powder and with sulfur powder located upstream. In marked contrast to previous work[20], we did not use seeds to nucleate growth; instead, we found that the best growth was obtained with carefully-cleaned substrates (see supplemental Figure S1). Because of the inherent variability of solid-source CVD, the growth process differed from run to run, resulting in average island sizes ranging from 1 μm to 100 μm. As detailed below, optical and structural characterization confirmed that the islands were predominantly composed of highly crystalline single $MoS_2$ layers; however, as shown in Figure 1(a), we frequently found very small regions of bi-, or multilayer $MoS_2$ at the centers of the crystals.

Photoluminescence (PL) is a valuable tool for probing the electronic structure and quality of $MoS_2$.[18] Figure 1(b) shows photoluminescence spectra from monolayer and bi-layer regions on a single CVD-grown island. The narrow spikes at high energy are the Raman transitions, detailed further in Supplementary Figure 2. The strong

PL peak at 1.84 eV and weaker peak at 1.95 eV match the A and B direct gap optical transitions observed in exfoliated monolayers of $MoS_2$. Likewise, the intensity of the bi-layer peak is much weaker (~7%) than the monolayer peak, reflecting the expected evolution from a direct-gap to an indirect-gap semiconductor. While the indirect bandgap transition is not visible in the bi-layer, this is not unusual on a $SiO_2$ substrate. Surprisingly, the measured peak width of 50-60 meV is similar to that observed for freely-suspended samples of exfoliated $MoS_2$ (50-60 meV) and is much narrower than that of $MoS_2$ exfoliated directly on oxide (100-150 meV)[18]. These results suggest that the CVD-grown samples are either of higher quality than exfoliated samples or in a cleaner electrostatic environment as a result of the growth process.

To characterize the crystal structure of these islands, we used a combination of atomic-resolution scanning transmission electron microscopy (STEM) imaging and lower-resolution electron diffraction techniques. For TEM imaging, the $MoS_2$ islands were transferred to thin-film silicon nitride or Quantifoil grids (See methods)[3]. Figure 1(c) shows an image of the atomic structure of the CVD $MoS_2$ monolayers taken by aberration-corrected annular dark-field scanning transmission electron microscopy (ADF-STEM). In ADF-STEM, a 60 keV, Angstrom-scale electron beam scans over the sample, and an image is formed by collecting the medium- to high-angle scattered electrons. Properly calibrated, ADF-STEM images each atom with contrast that varies roughly as the square of the atomic number Z.[29] Thus, the brightest spots in Figure 1(c) are the molybdenum atoms and the dimmer spots are the two stacked sulfur atoms. The hexagonal lattice is clearly visible, as indicated by the overlaid schematic representation and the structural model inset in Figure 1(c). No defects, atomic substitutions, or voids were initially observable within a single crystal. We note, however, that although the beam energy is below the ~80 keV threshold for damage in $MoS_2$[30], defects do readily form in the $MoS_2$ lattice under extended imaging, a process possibly catalyzed by surface contaminants from the transfer process.

We used selected-area electron diffraction (SAED) and dark-field TEM (DF-TEM), which maps the lattice information in the diffraction pattern onto real-space, to characterize the large-scale crystal structure of the samples[3,31]. Figure 1(d) shows a DF-TEM image and SAED pattern (inset) of a single triangular island roughly 45 μm across. The ~2-4 μm brighter and darker areas are bi-layer $MoS_2$[32]. The 6-fold symmetry in the position of the diffraction spots demonstrates that the triangle is a single crystal, with no rotational boundaries; the relative intensities of the spots indicate that the region is a single layer of $MoS_2$ in the 2H phase[33]. Analysis of dozens of triangles in several different growth runs shows that triangular-shaped islands are predominantly single crystals. Under continued growth, these islands merge together to form dense aggregates or continuous sheets (Supplementary Figure S3).

We further use electron diffraction techniques to uniquely identify the crystal orientation and analyze the edge structure of the $MoS_2$ islands[33]. As seen in the atomic resolution image of Figure 1(c), the lattice of monolayer $MoS_2$ exhibits only 3-fold symmetry (in contrast to the 6-fold symmetry of graphene). In particular, the real-space lattice lacks mirror symmetry across the [-1010] or zig-zag direction. This broken symmetry manifests itself directly in the diffraction data. As shown in Figure 2(a), there is a 10 % difference in the relative intensities of the [10-10] and [-1010] spots for the triangle shown in Figure 2(b). We calculated the diffraction pattern for a monolayer of $MoS_2$ using Bloch wave simulations to account for the complex scattering that allows the breaking of Friedel's rule and produces this asymmetry (See Supplementary Figure S4).[34] We also double checked with high-resolution imaging to confirm that the asymmetric diffraction pattern in Fig 2(a) corresponds to the indicated orientation of the $MoS_2$ lattice. Specifically, each of the brighter [-1010] spots point towards the Mo atom sites in the real space lattice. We can thus directly use the asymmetric diffraction pattern to uniquely identify the lattice orientation of $MoS_2$ crystals. Figure 2(c) shows a single dark-field TEM image taken of two nearby triangles that are nearly anti-aligned with one another. The intensity of the two triangles differs by 10% —the real space manifestation of the asymmetric diffraction pattern produced by each crystal.

Figures 2(b) and 2(c) show two different morphologies of $MoS_2$ triangles, with their crystal orientations overlaid. These examples, produced in separate growth runs, represent the two dominant crystal orientations found. In Figure 2(b), the triangle is oriented along the molybdenum zig-zag (Mo-z-z) edge, while in Figure 2(c), the triangle is oriented along the sulfur zig-zag (S-z-z) edge. These two crystal orientations are commonly observed in MoS$_2$ nanocrystals[35], and they are the two most energetically stable edge orientations[36]. We note that these measurements only define the microscale scale orientation of the edges; they do not determine the precise edge termination or structure on the atomic scale. Indeed, high-resolution TEM images suggest that the edges are rough at the nanoscale.

We consistently observe that Mo-z-z triangles (Figure 2(b)) have sharper and straighter edges than S-z-z triangles (Figure 2(c)). This morphological difference allows us to rapidly identify the crystal edges and orientation of triangles on the growth substrate simply by optical microscopy. In doing so, we also observe that all crystals from the same growth run have the same morphology, i.e., triangles from a given run will either be dominated by Mo-z-z or by S-z-z edges, a preference we attribute to kinetic effects. Both triangle morphologies exhibit the same range of average sizes from 1-100 μm. These classifications are important for understanding growth dynamics and suggest the possibility of refined control of edge morphology by tuning the CVD process.

In addition to single-crystal triangles, the CVD process produces polycrystalline islands, which can be characterized in detail by dark-field TEM. When neighboring nucleation islands grow together, they form tilt boundaries. Figure 3(a) is an example of a tilt boundary formed by the intersection of two S-z-z triangles with a 45° angular mismatch. Figure 3(b) shows the corresponding dark-field overlay, constructed by false coloring and overlaying the two DF-TEM images taken for the two diffraction spots. The grains have grown together to form an abrupt, faceted tilt boundary. In some cases, we do not see just such highly symmetric triangles but rather, as in Figure 3(c), regions of many small jagged, disordered crystals. The inset diffraction pattern reveals an even distribution of grain orientations in this region, as expected for nucleation on the amorphous $SiO_2$ substrate. The corresponding dark-field overlay in Figure 3(d) shows that the disordered crystals are polycrystalline aggregates with faceted boundaries. As indicated by the arrows, the $MoS_2$ grains sometimes overlap to form bi-layer regions. These crystal interface types with abrupt and overlapped boundaries, have also been seen in CVD-grown graphene[3,5,37]. Unlike in graphene, the grain boundaries in $MoS_2$ tend to be strongly faceted on the micrometer scale.

In addition to tilt boundaries, we have also discovered an important and common line defect in $MoS_2$ crystals grown by CVD—mirror, or symmetric, twin boundaries. In monolayer $MoS_2$, mirror twins are the intersections of two $MoS_2$ crystals with a relative in-plane rotation of 60° (or 180) that effectively swaps the positions of Mo and S lattice sites. Figure 3(e) shows the bright-field image of two triangles meeting at 180° and Figure 3(f) shows the corresponding DF-TEM image, generated from a single diffraction spot. Because they are not accompanied by any change in position of the diffraction spots, mirror boundaries can be easily missed in a simple diffraction analysis. Similar to the two triangles shown in Figure 2(c), however, twinning appears clearly as an intensity change in DF-TEM images. Unlike the tilt boundaries, the contact twins in Figure 3(e-f) grow within nucleation islands.

Figures 3(g-h) and Supplementary Figure S5 show another commonly observed shape that contains mirror twin boundaries: a highly symmetric, six-pointed star containing cyclic twins in these islands most likely formed during growth. The six-pointed stars resemble crystal morphologies seen in other inorganic nanocrystal growths and geological crystals[38].

We next examine the atomic-level structure of $MoS_2$ grain boundaries. Figure 4(a) displays a mirror-twin grain boundary in a 6-pointed star, imaged at high resolution using ADF-STEM. While the grain boundary follows the zig-zag direction on the micrometer scale, it exhibits faceting of ±20° relative to this direction on the nanometer scale. This nanoscale faceting is surprising because it avoids the highly symmetric zig-zag boundary structure which would correspond to the microscale alignment seen in Figures 3 (f) and (h), . Figure 4(b) shows the atomic structure of the grain boundary from ADF-STEM. The overlaid purple and green polygons show the boundary is formed primarily from 4- and 8-membered rings, with a recurring periodic 8-4-4- ring motif. Similar, even-numbered ring sizes have been predicted to form grain boundaries in monolayer h-BN because they avoid homoelemental bonds[39]. In the 8-4-4 motif, neighboring 4-rings meet at a sulfur site, which appears to have 4 nearest neighbors rather than the usual three—a change in coordination that does not occur in the theoretically predicted armchair and zigzag boundary structures (Supplementary Figure 6). . The $MoS_2$ mirror twin boundary contrasts with grain boundaries in graphene, which are more commonly formed by 5- and 7-membered rings that do not alter carbon $sp^2$ bonding[3,40].

In order to predict the effect of grain boundaries on the structural and optoelectronic properties of MoS$_2$ monolayer membranes, we employ first principles density functional theory (DFT) to model the experimentally observed grain boundary[41]. Figure 4(c) shows the energy-minimized structure of the mirror grain boundary, confirming that the observed boundary is at least a local energetic minimum. The model shows that the neighboring 4-rings do not break symmetry by having the center top and bottom sulfurs bond to different neighbors or shift out of vertical registry. The model, however, does not include the possibility of other elemental species reacting with dangling bonds out of plane. Figure 4(d) displays the corresponding density of states (DOS) of monolayer MoS$_2$, a grain boundary implanted into monolayer MoS$_2$, and just the boundary alone. Figure 4(e), which shows the local DOS in the bandgap, confirms the spatially localized nature of these mid-gap states by calculating the local DOS integrated in a window around the Fermi energy and shows that the calculated gap structure is not an artifact of the model used for computation (see methods). Analogous calculations on alternative zig-zag and armchair grain boundaries (See Supplemental Figure S6) confirm that this behavior is generic. These localized mid-gap states, typical of defects in semiconductors, are important because they can affect the optical and transport properties of the material[4,42].

Using the new understanding of the grain structure developed above, we can now identify individual grain boundaries on the growth substrate to perform systematic exploration of the effects of grain structure on the physical, chemical, optical, and electronic properties of MoS$_2$ membranes. In the two examples below, we examine the influence of individual grain boundaries on the materials photoluminescence and electronic transport properties.

We measure the effect of the grain boundaries on the optical properties using photoluminescence mapping, where we step the focused excitation laser over the sample and record a PL spectrum at every point. Figure 5(a) shows optical images of a mirror twin crystal and a tilt boundary crystal, produced in different growths. Figures 5(b-d) are maps of the integrated photoluminescence intensity, the peak position, and the peak width, corresponding to the crystals in 5(a). Supplemental Figures S2(b-c) show the corresponding Raman maps of the tilt boundary. Away from the boundaries, the triangles in both crystals show similar emission intensities with peak positions of 1.84 eV and widths of 57 meV in the mirror crystal and 60 meV in the tilt crystal. In the mirror twin crystal, the boundary is clearly visible as a straight line with 50 % quenching of the photoluminescence intensity, combined with an 8 meV up-shift in peak energy and 5 meV peak broadening. In the tilt crystal, the boundary appears as a faceted line, similar to that shown in Figure 4(b), with a surprising 100 % enhancement in emission strength at the boundary, combined with a 26 meV up-shift and 5 meV broadening. Photoluminescence quenching commonly arises from defects in semiconductors, such as the predicted midgap states at the boundaries, which can act as centers for non-radiative recombination[43]. While the amount of material structurally modified by the boundary is small compared with the 500-nm laser spot size, the effect can be enhanced by the diffusion of photogenerated excitons to the boundary. We estimate this latter process using the carrier mobility determined below and the measured emission time of 40 ps in our samples[44] (Supplemental Figure S7). The resulting exciton diffusion length of 15-20 nm is insufficient to explain the strong modification of the PL by boundaries. A different possibility for the mechanism by which grain boundaries strongly modify the observed photoluminescence is through changes in the doping levels of the $MoS_2$ material in the vicinity of the boundary, which might result from the midgap states described above, as well as from influence on the CVD growth process. The charge density in $MoS_2$ has been shown to dramatically alter the emission properties through the shifts in the balance of neutral and charged excitons[45]. Importantly, such a mechanism could explain either a quenching or enhancement of the emission, depending on the change in sample doping. The highly sensitive PL probe might also be influenced by compressive strain that could either modify the bandgap[46] or, alternatively, cause the boundary region to lift off the electrically disordered surface, thus enhancing the PL emission[18]. In general, getting control of the quenching and enhancement at grain boundaries is important for fabricating $MoS_2$ base optical and photovoltaic devices.

To determine the effects of grain boundaries on electrical transport we fabricated field effect transistors (FETs) by electron-beam lithography in three different configurations: within a grain (pristine), across the grain boundary (perpendicular), and along the grain boundary (parallel). Figures 5(e) and (f) show transfer curves for four such devices measured at room temperature using the Si growth substrate as a back-gate and a source-drain bias of 500 mV. The pristine devices show *n*-type behavior with mobility of 3-4 cm$^2$V$^{-1}$s$^{-1}$. Measured mobilities are nearly identical within a single crystal, while mobilities for different crystals range from 3-8 cm$^2$V$^{-1}$s$^{-1}$, with on/off ratios of 10$^5$-10$^7$. These values are comparable to those reported for back-gated field effect transistors (FETs) fabricated with mechanically-exfoliated MoS$_2$ on SiO$_2$[24,26,47-49] (in the absence of high-K dielectrics) and equivalent to the best reported values for monolayer CVD MoS$_2$[20-22]. The perpendicular device shows nearly identical performance to the pristine ones, indicating that the grain boundary has little effect on channel conductivity. This stands in contrast to other materials such as complex oxides where a single grain boundary can lead to a million-fold increase in resistance, or CVD-grown graphene, where both very high resistance or low resistance boundaries can significantly alter device characteristics[3,5,10].

The parallel device also exhibits similar behavior as well, but with 25% larger on-state and 60% larger off-state conductivity compared with the pristine devices. This pattern was repeated for four parallel devices fabricated on two flakes, although with considerable variability in the degree of conductivity increase. The relatively larger parallel conductivity is consistent with a picture of conduction through mid-gap states (discussed above) at the grain boundary. As a rough estimate, our data impose an upper limit for conductivity along the one-dimensional grain boundary of 1.5 μS-μm in the on state ($V_g$=+70 V) and of 120 pS-μm in the off state ($V_g$=-70 V). These values are to be compared to overall sheet conductances for pristine devices of 2.2 μS/□ in the on state and 56 pS/□ in the off state.   This analysis indicates that the few-atom-wide grain boundaries, while still semiconducting, have similar conductivity to a 1 μm wide strip of pristine material. This result is consistent with the predicted presence of midgap states in the material. However, the full effect is probably limited by the disorder in the grain boundary which will make it impossible to have a continuum of states . While the grain boundaries only slightly increase the in-plane conductivity, they should be more important when engineering large-area heterostructures such as $MoS_2$-based solar cells, where the defects can lead to interlayer tunneling.

The work presented here is an important step to incorporating molybdenum disulfide into two-dimensional electronics.  We were able to produce monolayers of large-area, highly crystalline $MoS_2$ on insulating surfaces.  The highly crystalline structures allowed us to perform a systematic study of the crystal edges, grain structure, and grain boundaries by electron microscopy. By knowing the grain structure we were able to identify crystal orientation and location of individual grain boundaries on the surface and systematically compare the optical and electronic properties to electronic structure calculations. The combinations of electron microscopy, optical spectroscopy and electronic transport show that grain boundaries in molybdenum disulfide play an important role on the optical properties and slightly increase the in-plane electrical conductivity of this two-dimensional material.

MoS$_2$ is just the first member of a large family of layered transition metal dichalcogenides to be synthesized as a monolayer[50], so the techniques and lessons developed in this work extend beyond this specific material system. All the two-dimensional transition-metal dichacolgenides are expected to exhibit similar grain boundaries and structure. The above techniques are thus crucial for exploring synthesis strategies to optimize the grain properties for all the layered transition-metal dichalcogenides. The methods also provide a template for studies of the microscopic and macroscopic impact of grain structure on monolayer membranes, which will be distinct for each structure. Our results are thus a significant step forward in realizing the ultimate promise of atomic membranes in electronic, mechanical and energy harvesting devices.

**Methods(800 words):**

**Growth procedure**

We grew monolayer MoS$_2$ by chemical vapor deposition using a method similar to that reported in ref[20]. Growth substrates were Si with 285 nm of thermally-grown SiO$_2$. Substrates were cleaned in acetone and isopropanol, followed by two hours in H$_2$SO$_4$/H$_2$O$_2$ (3:1) and 5 minutes of O$_2$ plasma. They were then loaded into a 2 inch CVD furnace and placed face-down above a crucible containing 14 mg of MoO$_3$ (≥99.5% Sigma Aldrich #100932642) with another crucible containing 120 mg of sulfur (≥99.5% Sigma Aldrich #101144903) located upstream.

The CVD growth is performed at atmospheric pressure while flowing only UHP nitrogen. The growth recipe is as follows: Sit 4 hours at 105 C with 500 sccm, ramp to 700 C at 15 C/min with 10 sccm, sit 5 minutes at 700 C, Cool to 570 C without feedback (~20 minutes) with 10 sccm, open furnace and flow 500 sccm for rapid cooling.

**TEM sample preparation:**

For the TEM grids, we 10 nm thick amorphous nitride windows (TEMwindows.com) for DF-TEM samples and holey Quantifoil grids (Ted Pella) for ADF-STEM. Polymer transfer layers were prepared by spinning PMMA A2 onto finished $MoS_2$ on silicon oxide/silicon chips at 4000 RPM for 60 seconds. The chips were floated on the surface of 1M KOH. The KOH etched the silicon oxide epi-layer, causing the chips to fall off and leaving the polymer and $MoS_2$ coated polymer membrane floating on the surface. The membrane was scooped out and transferred to DI water several times, then scooped up by a TEM grid and dried. The PMMA was removed by baking the TEM grids in a UHV heater at 300 C for 10 minutes or an atmospheric pressure $Ar/H_2$ gas flow for 4 hours. Before atomic-resolution imaging, we baked the samples overnight in UHV at 130 C.

**ADF-STEM**

ADF-STEM imaging was conducted using a NION UltraSTEM100 operated at 60 kV. Imaging conditions were similar to those used in[29]. Using a 25–30-mrad convergence angle, our probe size was close to 1.3 Å. Images presented in Figure 1 and 4 were acquired with the medium-angle annular dark-field detector with acquisition times of between 16 and 40 μs per pixel. The atomic resolution images in Figure 4 were smoothed to improve contrast.

**DF-TEM**

TEM imaging and diffraction were conducted using a FEI Technai T12 operated at 80 kV, which did not cause any apparent damage to the $MoS_2$ membranes at these lower dose densities. Acquisition time for dark-field TEM images were 5–50 s per frame. We used both displaced-aperture and centered DF-TEM for the images in the main text.

**DFT Modeling:**

Density functional theory calculations were performed with the PW91 generalized gradient approximation for the exchange-correlation functional and ultrasoft pseudopotentials, as implemented in the Quantum Espresso electronic structure package[41]. Structural relaxations were carried out at the gamma point and single-point calculations were done with appropriately converged Monkhorst-type k-point grids. Supercells were built with about 10 Å separation between replicas to achieve negligible interaction.

**Photoluminescence:**

The photoluminescence measurements were performed in a Renishaw InVia Raman Microscope using a 532 nm laser and a 1800 l/mm grating. The curves in Figure 1 were taken at 100 µW laser power for 10 seconds. The maps were taken at 100 µW laser power for 1 second with 300 nm step size.

**Electrical measurements:**

We fabricated back-gated FETs directly on the growth substrates. Grain boundaries were identified by photoluminescence (PL) as shown in Figure 4. The $MoS_2$ was shaped by patterning a PMMA mask with ebeam lithography and etching with $O_2$ plasma for 20 seconds at 50 W with 20 sccm flow rate. The electrodes were then patterned by electron beam lithography with a bi-layer PMMA stack and a subsequent evaporation of Al/Cr/Au (50 nm/5 nm/50 nm). We did not anneal the samples after fabrication.

All devices were measured at room temperature under $10^{-5}$ Torr inside a lakeshore probe station. We used an Agilent 4155C Semiconductor Parameter Analyzer to perform 2-point conductance measurements. FET transfer curves shown in Figure 4 were taken at 500 mV drain bias. The devices showed ohmic contacts from -500 mV to 500 mV source-drain bias.

**Contributions:**

A.M.v.d.Z supervised and coordinated all aspects of the project. $MoS_2$ growth was carried out D.A.C. and A.M.v.d.Z. Electrical characterization and analysis was carried out by A.M.v.d.Z., D.A.C. and G.H.L. under the supervision of J.C.H. Electron microscopy and data analysis were carried out by P.Y.H. with D.A.M.'s supervision.

Optical spectroscopy and data analysis were carried out by A.M.v.d.Z. and Y.Y. under T.F.H.'s supervision. DFT simulations were carried out by T.C.B. under D.R.R.'s supervision. A.M.v.d.Z., P.Y.H., D.A.C., T.C.B., Y.Y., T.F.H., D.A.M., and J.C.H. wrote the paper.

**Acknowledgements:**

Overall project coordination, sample growth, electrical and optical characterization were supported as part of the Center for Re-Defining Photovoltaic Efficiency Through Molecular-Scale Control, an Energy Frontier Research Center funded by the U.S. Department of Energy (DOE), Office of Science, Office of Basic Energy Sciences under Award DE-SC0001085. A.M.v.d.Z was supported by the EFRC as a research fellow. Electron microscopy was performed at and supported by the Cornell Center for Materials Research, an NSF MRSEC (NSF DMR-1120296). P.Y.H. was supported under the National Science Foundation Graduate Research Fellowship Grant No. DGE-0707428. D.A.C. was supported by a Columbia University Presidential fellowship and a GEM Ph.D Fellowship sponsored by the Center for Functional Nanomaterials at Brookhaven National Lab. T.C.B. was supported under the Department of Energy Office of Science Graduate Fellowship Program (DOE SCGF), administered by ORISE-ORAU under Contract No. DE-AC05-06OR23100. The authors thank Sasha Gondarenko, Inanc Meric, Jayakanth Ravichandran, Lei Wang, Justin Richmond-Decker, and Philip Kim for helpful discussions.

**Figure Captions:**

**Figure 1: Growing single-crystal MoS$_2$.** a) Optical reflection image of a monolayer MoS2 triangle grown by CVD on a 285 nm silicon oxide thin film on silicon substrate. The triangle is 123 µm from tip to tip. b) Photoluminescence spectra from monolayer and bi-layer MoS$_2$. Peak height is normalized to the silicon Raman peak. The narrow spikes at high energy are the Raman transitions which are discussed in Supplementary Figure S2(a). c) High resolution ADF-STEM image of a freely suspended, monolayer MoS$_2$ on a TEM grid with the schematic representation overlaid and the structural model inset. The bright spots are molybdenum atoms; the grey spots are two stacked sulfur atoms. The lattice is composed of hexagonal rings alternating molybdenum and sulfur sites.
d) Dark-field TEM image of a large triangle on an amorphous nitride TEM grid with the diffraction pattern inset. The 6-fold diffraction spots show the triangle is a single crystal, and the dark-field image shows that the crystal is continuous. The bright and dark patches in the image are oriented bi-layers.

**Figure 2: Crystal orientation and edge terminations.** a) Diffraction pattern from a single triangle with a line profile take through the diffraction spots. The profile shows a 10% asymmetry between the [-1010] spot and the [10-10] spots. We use the asymmetry to infer the real space lattice orientation, overlaid on the diffraction pattern. b) Bright-field image of the triangle corresponding with the diffraction pattern shown in a), with the crystal lattice overlaid. The sharp straight crystal edges correspond with a Mo-zig-zag edge orientation. c) Dark-field image taken from a single diffraction spot show two triangles, each with different scattering

intensities. The intensity difference is the real space manifestation of the asymmetry in the single-crystal diffraction pattern in (a) and shows that the two triangles have nearly opposite lattice orientation. The curved crystal edges in both triangles correspond with a S-zig-zag edge orientation.

**Figure 3: Tilt and mirror twin grain structures.** a) Bright-field TEM image of two triangles that have grown together. Inset diffraction pattern shows the two crystal orientations are 45° apart. b) Color-coded overlay of dark-field TEM images corresponding with the red- and cyan-circled two spots in (a) show a tilt grain boundary as a faceted line connecting the two triangles. c) Bright-field image of a region containing multiple irregularly shaped $MoS_2$ islands. Inset shows multiple crystal orientations. d) Dark-field TEM overlay, color-coded by crystal orientation, and corresponding to (c) show that the irregular shapes are made up of polycrystalline aggregates. The crystals are connected both by faceted, abrupt grain boundaries, and ~ 1 µm overlapped bi-layers. e) Bright-field image of a contact twin composed of opposing triangles, nucleating from a single point with diffraction pattern inset. f) Dark-field image corresponding with the orange circle in (e) shows the two triangles have different diffraction intensity. This difference is the real space manifestation of the asymmetry in a single crystal diffraction shown in 2a and show that this is a mirror twin. g) Bright-field image of a 6 pointed star. Inset diffraction shows star has no rotational boundaries. h) Dark-field TEM image corresponding with the orange circle in (g) shows that the star contains several rotationally symmetric mirror crystals, forming a cyclic twin.

**Figure 4: Grain boundary atomic structure.** a) High resolution ADF-STEM image of a mirror twin boundary. The boundary is visible just below the annotated line. The annotation indicates the faceting of the boundary at +/- 20° off of the zig-zag direction at the nanoscale. b) Zoomed-in image of the grain boundary shows a periodic line of 8-4-4 ring defects. c) Energy minimization with DFT confirms that the boundary is locally stable. d) The total density of states (DOS) of pristine $MoS_2$ (black), the DOS of $MoS_2$ with the grain boundary (red dashed), and the DOS projected onto only those atoms along the grain boundary (blue filled). The dashed grey line denotes the Fermi energy of neat $MoS_2$ and the shaded area indicates the band gap. All states have been given a Gaussian broadening of 0.07 eV. e) A two-dimensional slice through the local DOS (the projection of the total DOS onto the individual atomic orbitals) integrated over a 1.7 eV range about the Fermi energy of the pristine $MoS_2$. The color scale of the density is 0-0.05 bohr$^{-3}$.

**Figure 5: Optical and electronic properties**. a) Optical images of a mirror twin crystal and a tilt crystal. b-d) are color plots of photoluminescence maps, where: b) Red is the relative quantum yield, with color scale 0-1100 a.u. We see 50 % quenching at the mirror boundary and a 100 % enhancement at the tilt boundary. Faceting similar to that seen in Figure 3b is visible along the tilt boundary. c) Green is the peak position, with color scale 1.82-1.87 eV. There is an upshift of 8 meV at the mirror boundary, and a much stronger 26 meV upshift in the tilt boundary. d) Cyan is the peak width with color scale of 53-65 meV. The peak broadens from 55

meV to 62 meV at the boundary in both samples. e-f) Linear and logarithmic electrical transport transfer curves of 4 FETs fabricated from the single mirror twin MoS$_2$ crystal shown in the inset of (e). The curves correspond with pristine crystals (magenta and cyan), and crystals containing a grain boundary running perpendicular (black) and parallel (orange) to the flow of electrons.

**Figure 1: Growing single-crystal MoS$_2$**

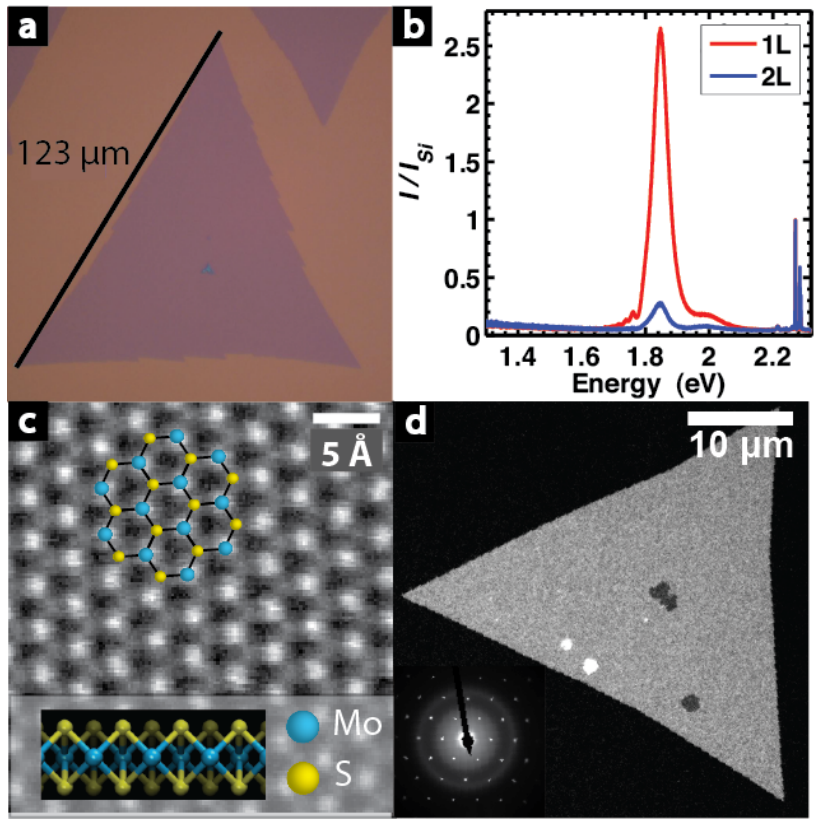

**Figure 2: Crystal orientation and edge terminations.**

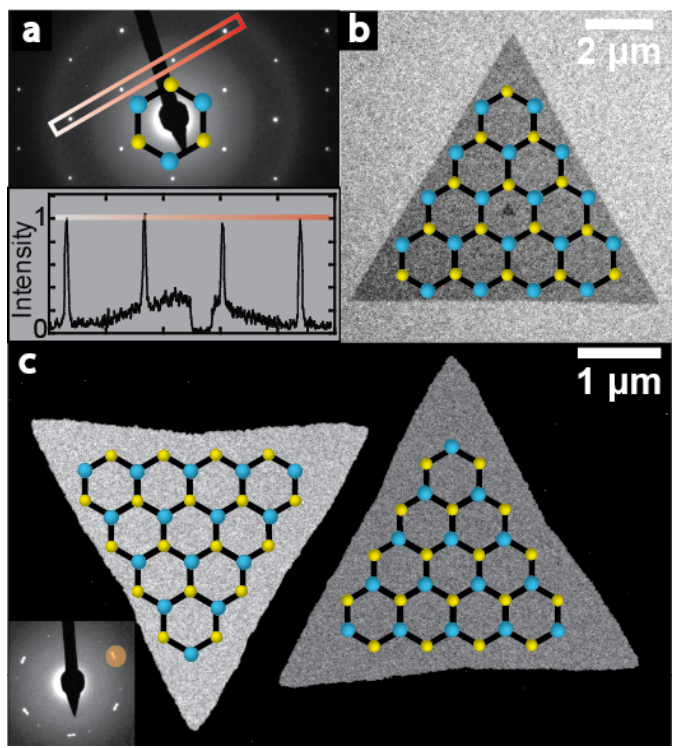

**Figure 3: Tilt and mirror twin grain structures**

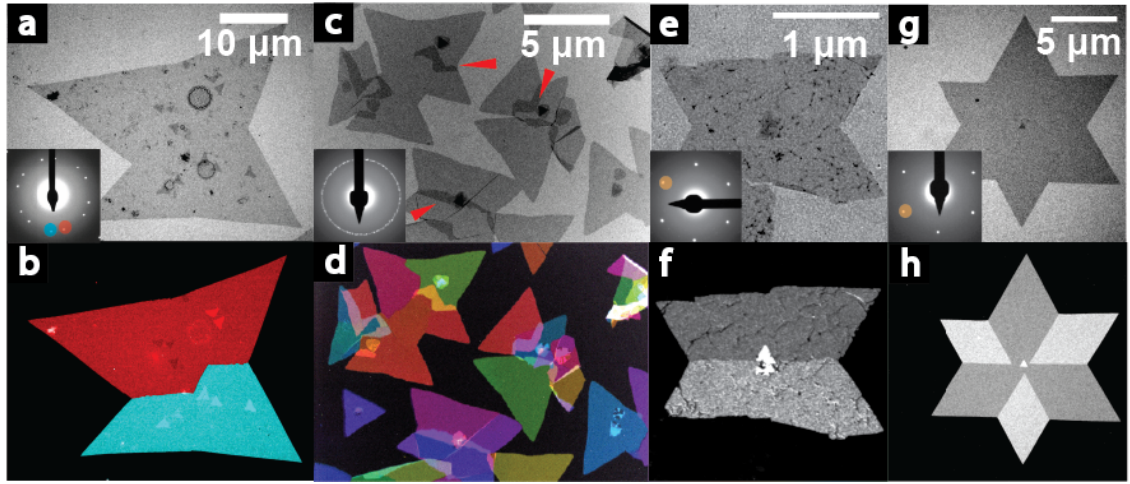

**Figure 4: Grain boundary atomic structure**

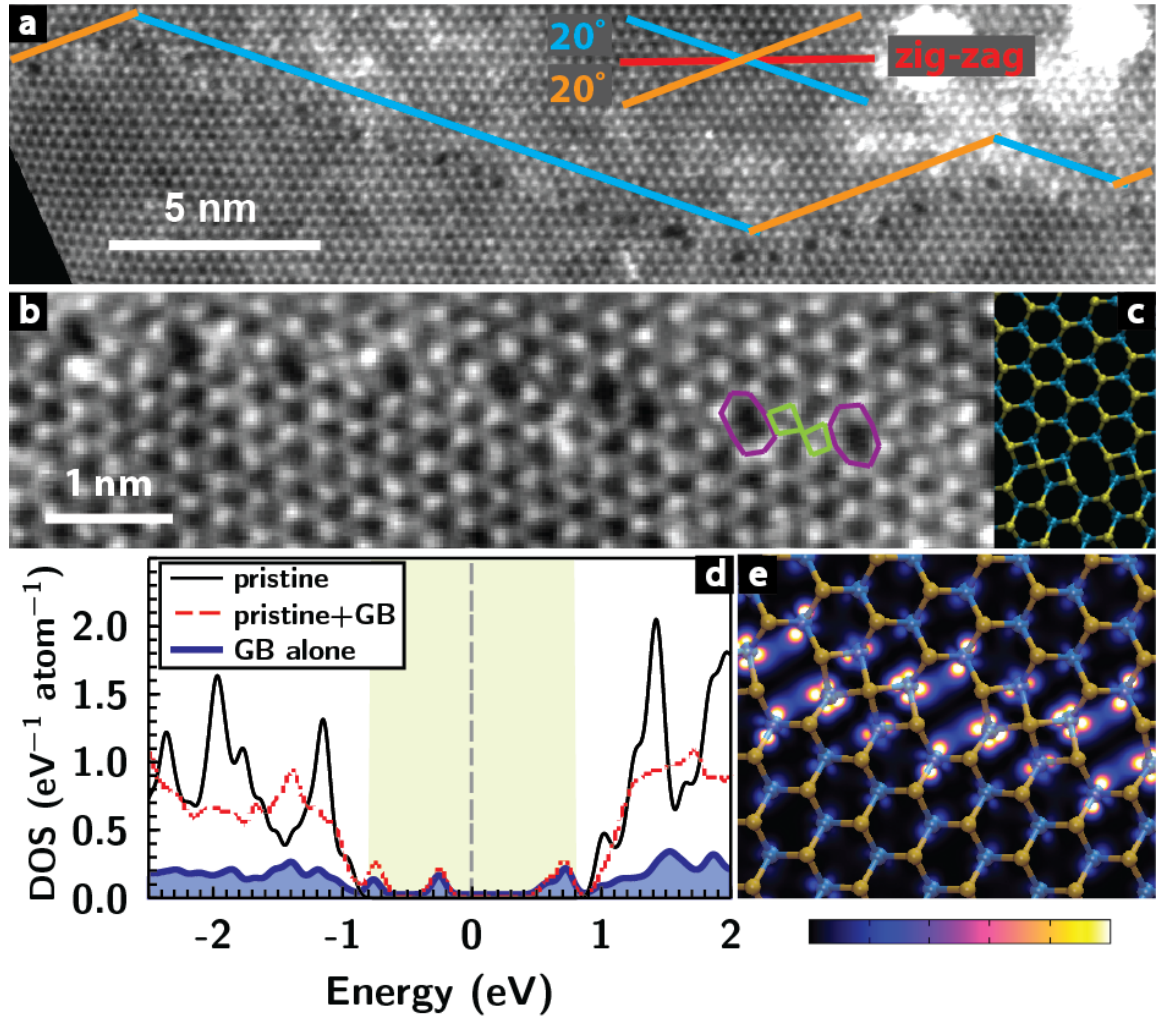

**Figure 5: Optical and electronic properties**

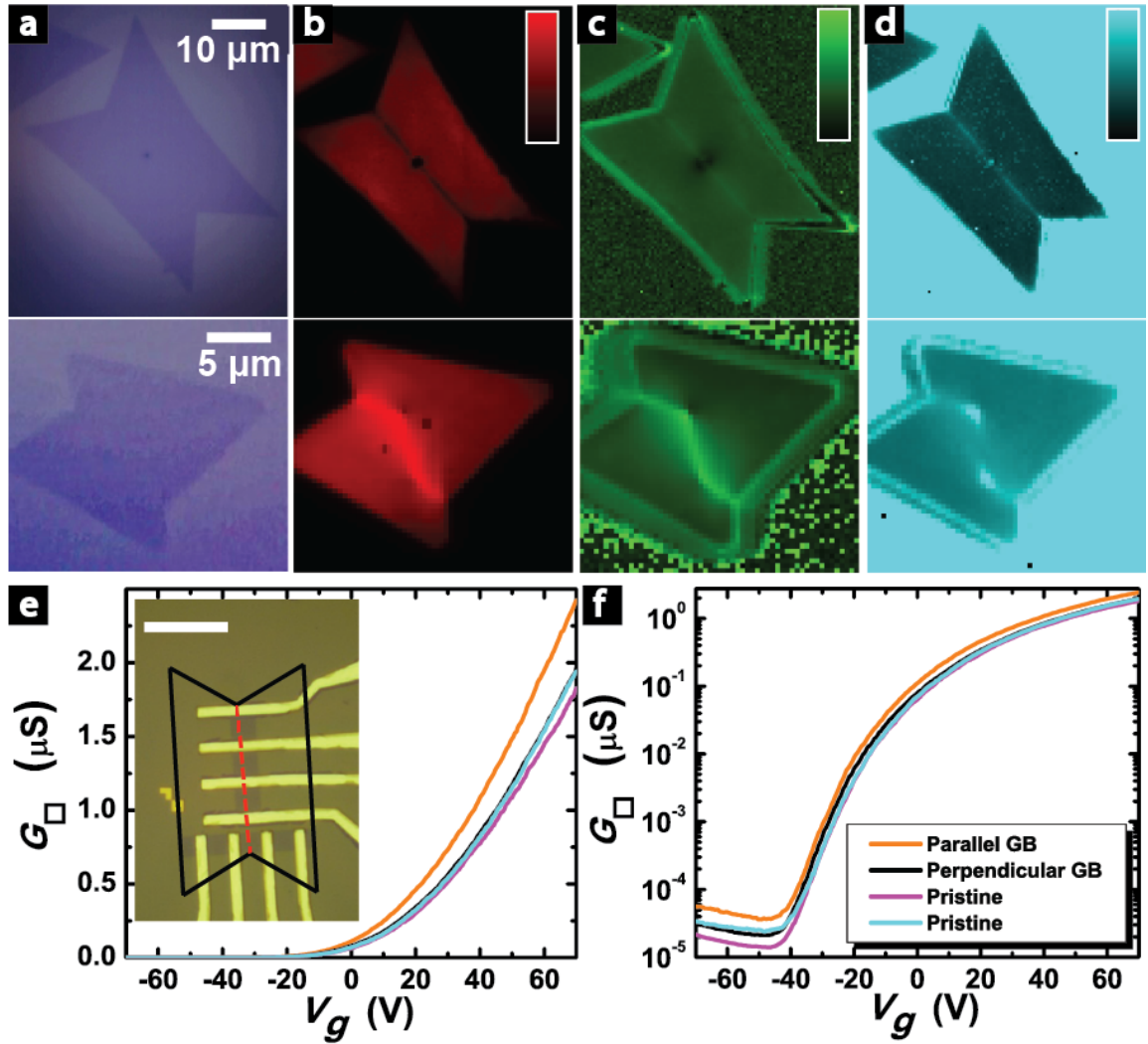

**Supplementary Materials:**

**Grains and grain boundaries in highly-crystalline monolayer molybdenum disulfide**

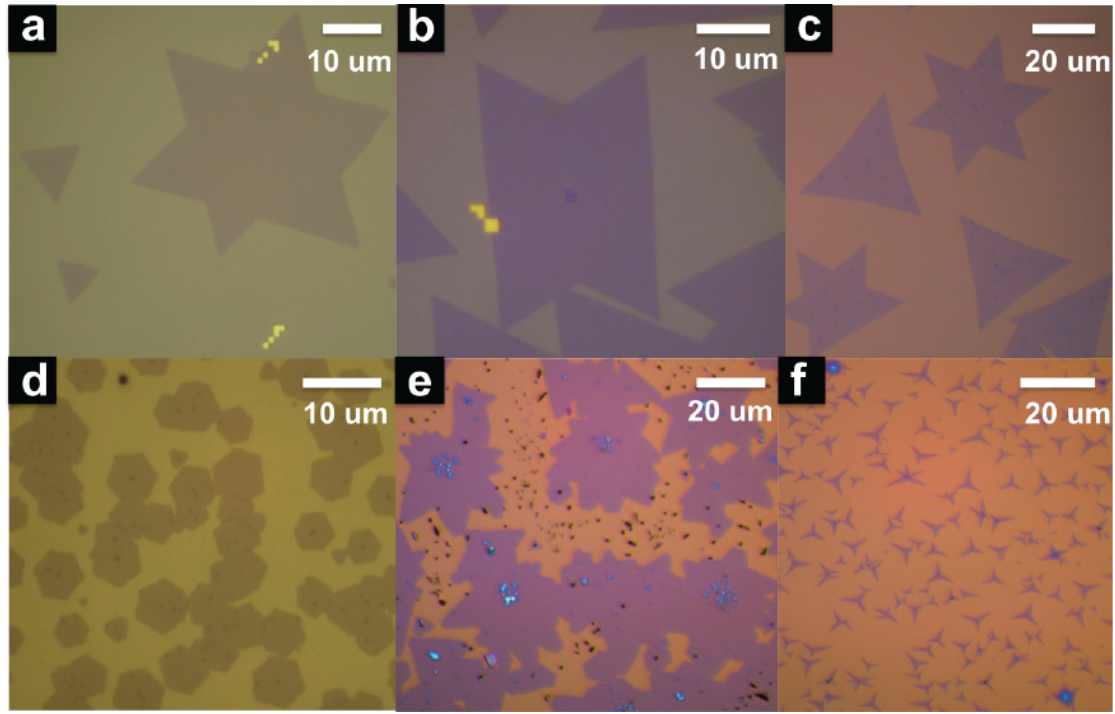

**Supplementary Figure S1: Commonly-observed shapes in MoS$_2$ CVD.**

Optical micrographs of various CVD MoS2 crystal shapes from different growths. a) Mo-z-z triangles and 6-point star grown on "clean" Si/SiO2 substrate. Note: metal alignment marks were deposited *after* growth. b) Mo-z-z mirror twin crystal used for electrical devices in Figures 5e-f. c) S-z-z triangles and 5- and 6-point stars. d) Hexagons. e) Gear-like polycrystalline structures grown on "dirty" substrates. f) 3-point stars grown on "dirty" substrates.

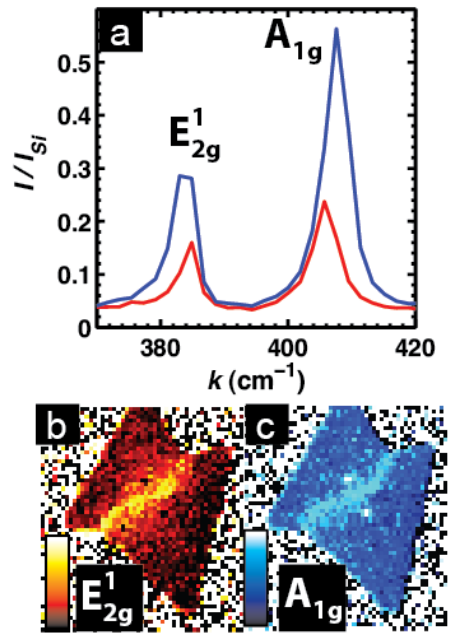

**Supplementary Figure S2: Raman spectra and mapping**

a) Raman spectra for monolayer and bilayer patches of $MoS_2$ corresponding with the spectra from Figure 1b. b-c) Raman maps of the Tilt boundary from Figure 4 for the b) $E_{2g}^1$ mode. Range 382.0 - 383.6 cm$^{-1}$. c) $A_{1g}$ mode. Range 402.9-404.2 cm$^{-1}$ Both modes show an upshift of 1 cm$^{-1}$ at the grain boundary. While this shift may indicate a change in strain or doping at the boundary, it is difficult to interpret compared with the more marked changes to the photoluminescence seen in Figure 5 of the main text.

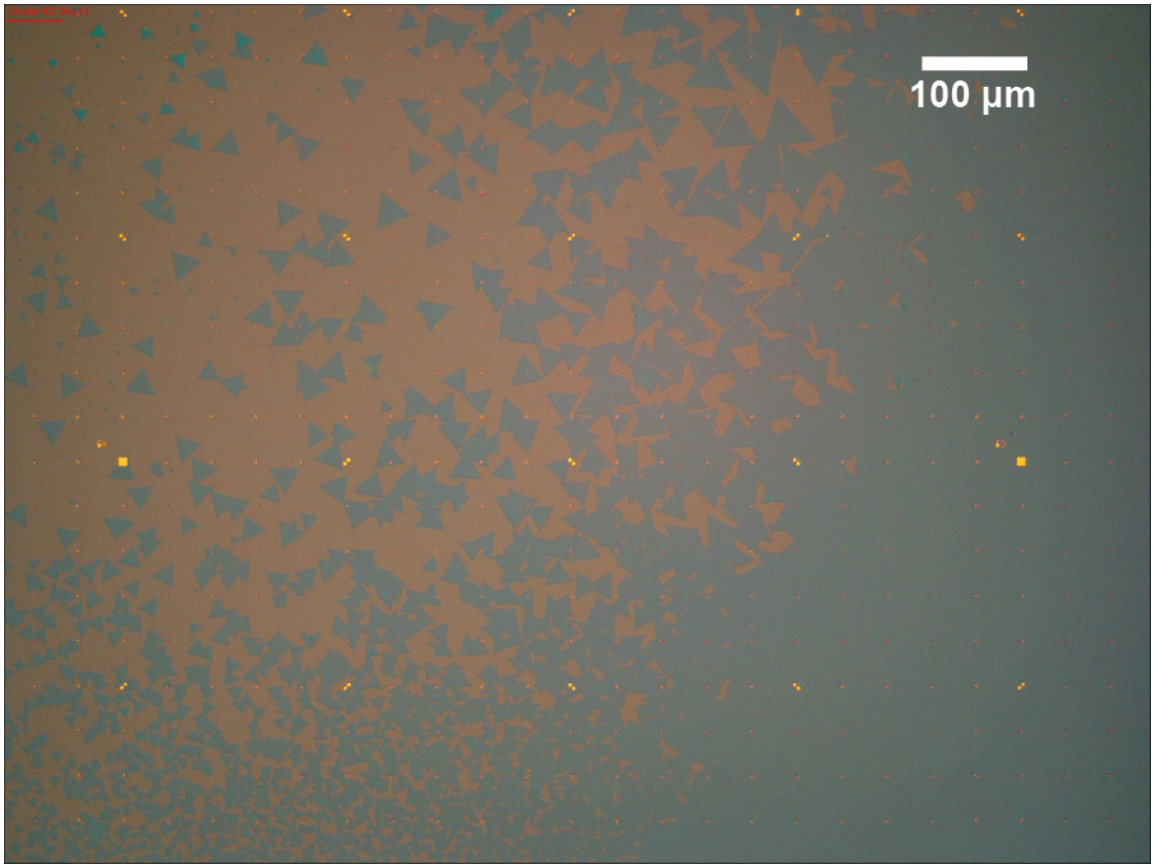

**Supplementary Figure S3: Continuous sheet**

CVD MoS$_2$ crystals can grow together to form continuous monolayer sheets. Top left is bare oxide with sparse crystals. Bottom right is continuous monolayer MoS$_2$. Alignment marks are placed after growth.

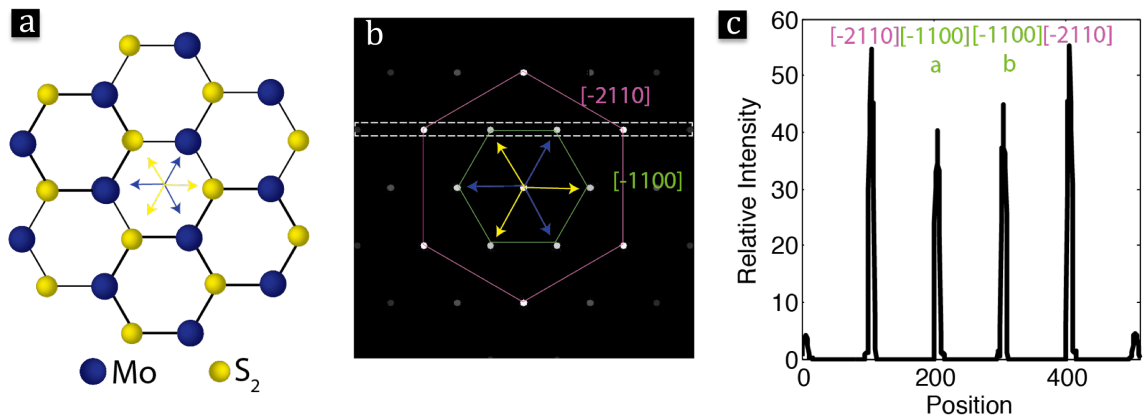

**Supplementary Figure S4: Bloch wave simulation**

a) Real-space lattice orientation of monolayer MoS2. b) Electron diffraction pattern for a monolayer of $MoS_2$ calculated using a Bloch wave CBED (convergent beam electron diffraction) code [1]. c) A line profile of the diffraction spots. The integrated intensity of three alternating [-1100] spots (blue arrows in b ) is 10% higher than the other three (yellow arrows) spots. This asymmetry occurs in the [-1100] spots, but not in the [-2110] spots. When the real and reciprocal lattices are placed side-by-side, vectors toward the "bright" diffraction spots in reciprocal space also indicate the vector from the center of a hexagon to a Mo site.

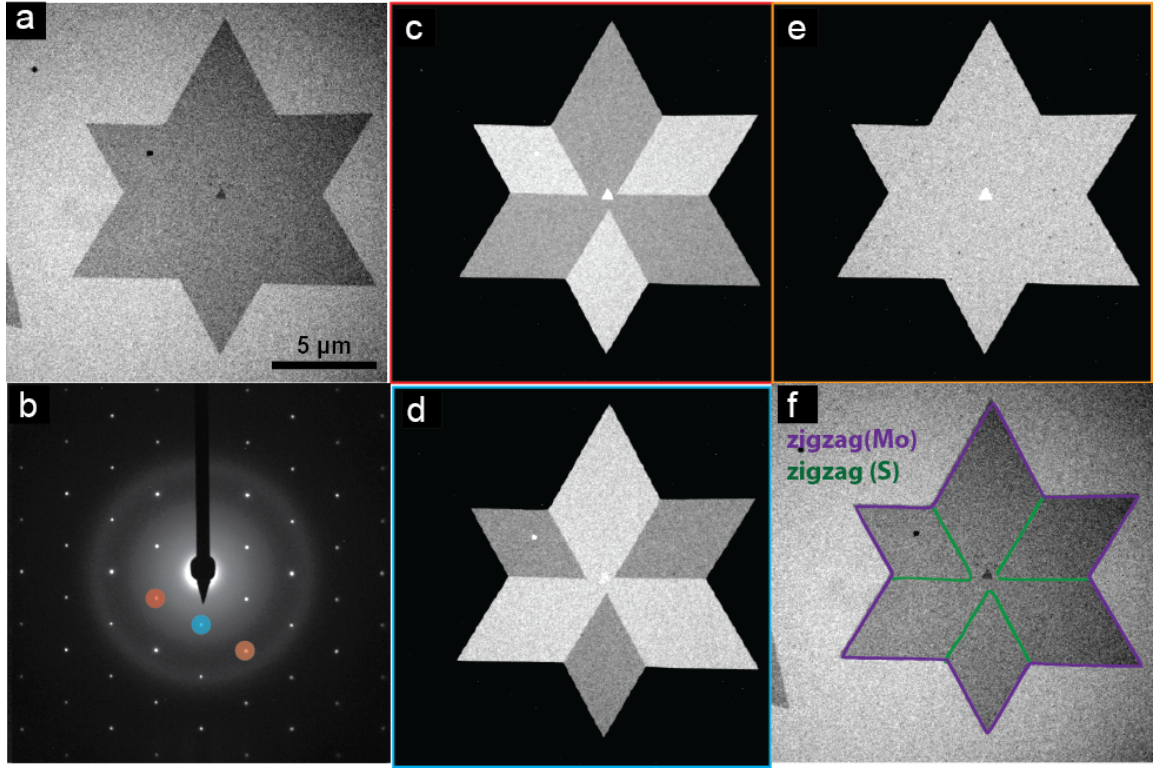

**Supplementary Figure S5: DF-TEM imaging of cyclic twin**

a) Bright-field image of a 6 pointed star. b) Full diffraction pattern shows star has no rotational boundaries. c-e) Dark-field TEM images corresponding with the c) red (-1010), d) blue (10-10), e) orange (-2110) spots in the diffraction image. The red and blue spots show opposite intensity due to swapping the in-phase and out of phase high intensity direction of scattering off the lattice. The orange spot shows an even intensity over the entire star corresponding with the symmetric diffraction pattern for a (-2110) spot in a single crystal. f) Bright-field TEM image with edge orientations extracted from Dark-Field measurements overlaid. The outer edges are oriented along the Mo-z-z direction, which the grain boundaries are oriented along the S-z-z directions.

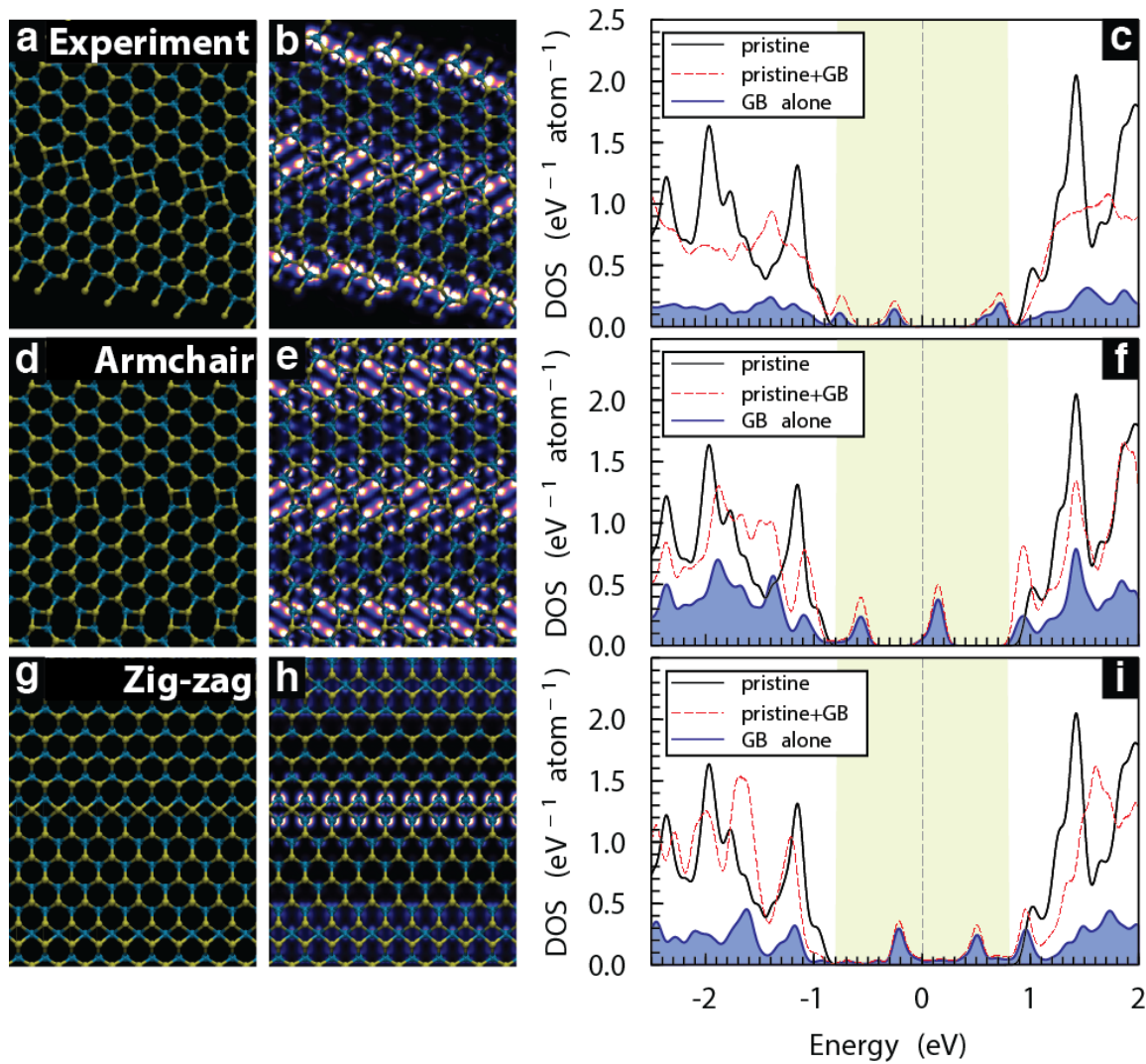

**Supplementary Figure S6: Simulating the grain boundaries**

a) Geometry optimized structure for the experimentally observed grain boundary, as determined by DFT (a). b) The local DOS (LDOS) integrated over a 1.7 eV window in the band gap of pristine $MoS_2$ c) confirming the spatial localization of mid-gap states, and the energy-resolved density of states including the projection onto the grain boundary atoms. In (c), we have subtracted off the contribution to the DOS arising from the unphysical edge atoms. Analogous calculations are shown for an armchair grain boundary d-f) and for a zig-zag grain boundary g-i), both which are entirely periodic in two dimensions. In panels (b), (e), and (h), the colorscale indicates the magnitude of the integrated LDOS, from 0 (dark) to 0.025 bohr$^{-3}$ (light).

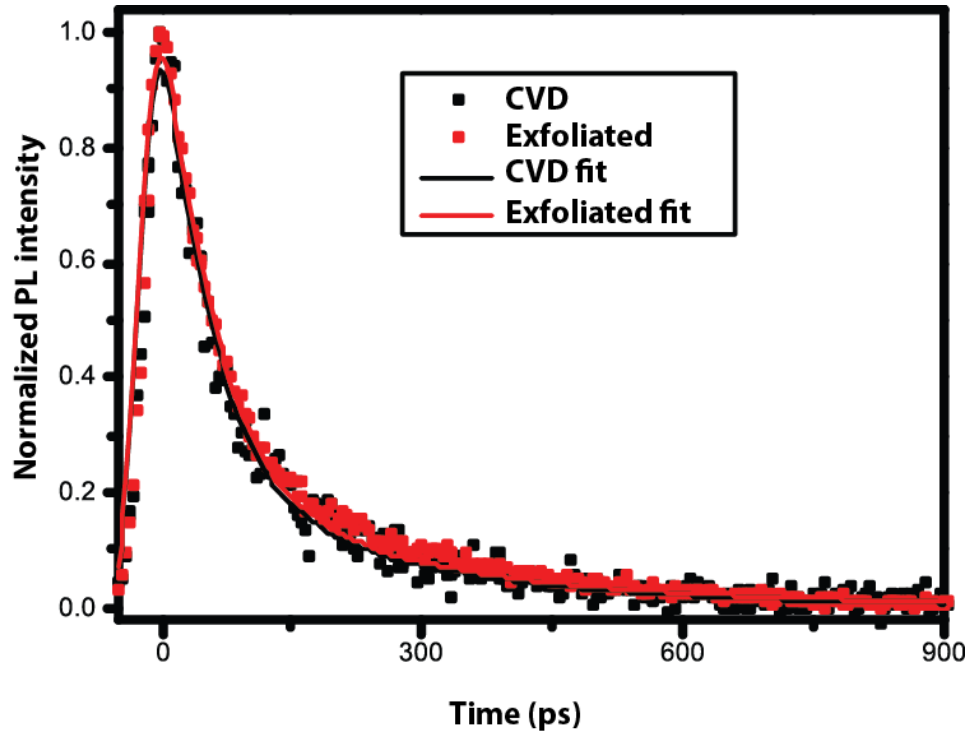

**Supplementary Figure S7: Time resolved photoluminescence**

Time-resolved photoluminescence measurements obtained by time-correlated single photon counting with femtosecond excitation by 400-nm laser pulses. Results for exfoliated and CVD MoS$_2$ samples on oxide layers yield, after accounting for the instrument response function, yield nearly identical time constants of $\tau_{exfoliated}$=44 ps, and $\tau_{CVD}$=42 ps.

**Supplementary Methods:**

**DFT calculations:**

Density functional theory (DFT) calculations were performed with the PW91 generalized gradient approximation for the exchange-correlation functional and ultrasoft pseudopotentials, as implemented in the Quantum Espresso electronic structure package[2]. Supercells were generated with about 10 Å separation to ensure negligible interactions between replicas. Structural relaxations were carried out at the gamma point until all components of all forces were less than 0.001 a.u. Pristine $MoS_2$ (3.12 and 2.32 Å for Mo-Mo and Mo-S bond lengths, respectively) energy calculations were done with a 16x16 Monkhorst-type k-point grid, confirming the material's direct band gap with a predicted energy of about 2 eV.

Because the direction of the experimentally observed 8-4-4 grain boundary is incommensurate with the periodicity of the underlying crystal, a system periodic in two dimensions cannot be constructed. Thus we employed the system shown in Figure S6(a) which is periodic along the direction of the grain boundary, but finite in the orthogonal direction such that the edge, terminated by S dimers, is about 10 Å away from the grain boundary. Energy calculations for this 87-atom supercell were performed with 5 k-points in the periodic direction. Local density of states (LDOS) analysis, shown in Figure S6(b) confirms that the electronic effects of the artificial edges are physically confined along the perimeter and so should not affect the properties along the grain boundary. These conclusions are also corroborated by a negligible change in bond length, compared to the bulk, for atoms away from the grain boundary. Analogous calculations on 2D periodic systems with pure armchair and zig-zag grain boundaries, shown in Figure S6 (d-i) similarly yield mid-gap states localized along the boundary, further indicating that the effect is generic and not an artifact of the finite strip size.

**Estimating non-radiative recombination:**

A factor that can contribute to strong contrast at grain boundaries in photoluminescence (PL) images is the diffusion of the photo-generated excitons to the boundary regions.[3] Such a process could provide a natural explanation for the strongly reduced PL observed from some boundaries in our samples, since non-radiative recombination may be more efficient in the presence of the localized structures and states of the boundary. Here we estimate the potential impact of this effect using measured values for the carrier mobility and exciton lifetime.

The electrical measurements in this paper yielded an electron mobility of $\mu_e = 3$ cm$^2$V$^{-1}$s$^{-1}$ for typical CVD-grown samples. From the Einstein relation for a temperature of $T = 300$ K, we then obtain an electron diffusivity of $D_e = \mu_e k_b T / e = 8 \times 10^{-2}$ cm$^2$s$^{-1}$. Since the electron and hole masses are similar, we assume that the exciton diffusivity is half of the electron diffusivity, *i.e.*, $D_{exc} = 4 \times 10^{-2}$ cm$^2$s$^{-1}$. The scattering mechanisms for electrons and excitons need not be the same, so this should only be considered as an estimate.

The characteristic length for diffusion of excitons towards the boundary from the 2-D bulk material over a time $\tau$ is given by $L = 2\sqrt{D_{exc}\tau}$. For a time interval of $\tau$ = 40 ps reported above for the exciton lifetime, we then infer a diffusion length of $L$ = 24 nm. This length, while not negligible, is small compared to the 500-nm spot size of the excitation laser. The quenching of the PL by 50% observed at some grain boundaries must consequently rely primarily, as discussed in the main text, on mechanisms other than exciton diffusion.

**Supplementary References:**